\documentclass[twocolumn,showpacs,preprintnumbers,amsmath,amssymb]{revtex4}

\def\ket#1{\mathinner{|{#1}\rangle}}
\def\braket#1{\mathinner{\langle{#1}\rangle}}

\usepackage{amssymb,graphicx}
\usepackage{dsfont}
\usepackage{subfigure}

\begin{document}

\title{Squeezing and photon counting with the cubic phase state}
\author{Seckin Sefi}
\email{seckin.sefi@mpl.mpg.de}
\affiliation{Institute of Physics, Johannes-Gutenberg Universit\"at Mainz, Staudingerweg 7, 55128 Mainz, Germany}
\affiliation{Optical Quantum Information Theory Group, Max Planck Institute for the Science of Light, G\"unther-Scharowsky-Str.1/Bau 26, 91058 Erlangen, Germany}

\begin{abstract}
Recently, a non-Gaussian state, which is called cubic phase state has been experimentally realized. In this work we show that, in case one has access to a proper cubic phase state, it is possible to make photon counting experiments and generate extremely squeezed states.
\end{abstract}

\maketitle

\section{Introduction}

For designing quantum optical systems, besides the linear optical tools and the Gaussian operations, most common non-Gaussian interaction that has been considered is the Kerr gate. Some of the things possible with the Kerr interaction are, creating superposition states such as the cat states \cite{Yurke1986}, \cite{Milburn1986}, \cite{Korolkova2001}, realizing controlled quantum gate for qubits \cite{PhysRevA.52.3489}, \cite{Turchette1995} and constructing photon detectors \cite{Kok2002}, \cite{Munro2005}. However, as far as we know, none of those proposals have been implemented experimentally because of the difficulties in realizing high amplitude Kerr interaction. 

On the other hand, recently, an approximation to a non-Gaussian state which is called cubic phase state: $\int e^{itx^3}\ket{x}dx$ \cite{Gottesman2001}, \cite{Ghose2007} has been realized \cite{Yukawa2013}. Yet, besides the works \cite{Sefi2011} and \cite{PhysRevA.88.012303}, which propose to use cubic phase state to generate Kerr interaction, it is not obvious in which tasks one can benefit from this state. Here we show that, using properly realized cubic phase state, one can obtain squeezed states that hasn't been possible with the conventional medium based methods and make photon counting experiments. The proposals presented in this work does not rely on any particular realization of the cubic phase state. Unfortunately the recently realized cubic phase state doesn't seem to be sufficiently ideal enough for the implementation of the proposals in this work because it is a first order approximation to the cubic phase state. A higher order realization or the proposal in ref. \cite{Gottesman2001} will be useful however. Nevertheless, exploring the opportunities of the cubic phase gate may lead to better realizations of the cubic phase state.

One important advantage of the cubic phase state is that it fits to the continuous variable gate teleportation model perfectly. This means that once the necessary state has been created it can teleported onto an arbitrary state deterministically. Thus it is possible to convert the cubic phase state into cubic phase gate: $e^{itX^3}$. For the continuous variable teleportation of the non-Gaussian gates please see the refs \cite{Menicucci2006}, \cite{Gu2009}, \cite{PhysRevA.88.012303}.


The article has been organized as follows, we first discuss squeezing with the cubic phase gate, then photon counting. Finally we discuss some issues and possible improvements regarding the implementation of the cubic phase gate.

Throughout, we use capital letters and hats for operators and small letters for scalars and functions. In equations, the letter $i$ is used only as the square root of -1. We use the convention $\hbar=1/2$, i.e., the fundamental commutation relation is $[X,P]=i/2$ with $X\equiv (\hat{a}^\dag+\hat{a})/2$ and $P\equiv i(\hat{a}^\dag-\hat{a})/2$.

\section{Squeezing}

Squeezing and squeezed states have great importance in many tasks ranging from gravitational wave detection \cite{Gea-Banacloche1987} to quantum information processing tasks \cite{Weedbrook2012}. In this section we will propose a squeezing operation using cubic phase gate, displacements and phase shifters.

Consider the following operator relations:

\begin{eqnarray}
e^{it_1P}e^{it_2X^3}e^{-it_1P}=e^{it_2X^3}e^{\frac{3}{2}it_1t_2X^2}e^{\frac{3}{4}it_1^2t_2X}e^{\frac{1}{4}it_1^3t_2}\\
e^{-it_1P}e^{-it_2X^3}e^{it_1P}=e^{-it_2X^3}e^{\frac{3}{2}it_1t_2X^2}e^{-\frac{3}{4}it_1^2t_2X}e^{\frac{1}{4}it_1^3t_2}
\end{eqnarray}

Combining these two equations we get:

\begin{equation}\label{eq:cubic_to_quadratic}
e^{it_1t_2X^2}=e^{it_1P}e^{i\frac{1}{3}t_2X^3}e^{-2it_1P}e^{-i\frac{1}{3}t_2X^3}e^{it_1P}e^{-\frac{1}{6}it_1^3t_2}
\end{equation}

Thus, we derived a formula for a second order operator using cubic phase gate and displacements. A similar formula has also been derived in ref. \cite{Sefi2011}. Now let us decompose this operator into phase operators and squeezing operator by using Bloch-Messiah decomposition \cite{Braunstein2005}. For this purpose we need to first find the mode transformation of this operator.

\begin{eqnarray}
e^{itX^2}\hat{a}e^{-itX^2}=\left(1-i\frac{t}{2}\right)\hat{a}-i\frac{t}{2}\hat{a}^\dag\\
=e^{i\phi_1}\sqrt{1+\frac{t^2}{4}}\hat{a}e^{i\phi_2}+e^{i\phi_1}\frac{t}{2}\hat{a}^\dag e^{-i\phi_2}
\end{eqnarray}

where:

\begin{equation}
\phi_1=-\frac{1}{2}\arctan{\frac{t}{2}}-\frac{\pi}{4}, \quad \phi_2=-\frac{1}{2}\arctan{\frac{t}{2}}+\frac{\pi}{4}
\end{equation}

This means that:

\begin{eqnarray}
e^{itX^2}=e^{i\phi_1\hat{a}^\dag\hat{a} }e^\frac{r(\hat{a}^2-\hat{a}^{\dag^2})}{2}e^{i\phi_2\hat{a}^\dag\hat{a}}\\
e^\frac{r(\hat{a}^2-\hat{a}^{\dag^2})}{2}=e^{-i\phi_1\hat{a}^\dag\hat{a}} e^{itX^2}e^{-i\phi_2\hat{a}^\dag\hat{a}}\label{eq:squeezing_to_quadratic}
\end{eqnarray}

and here the relation between squeezing parameter $r$ and $t$ is as follows:

\begin{equation}
\tanh{r}=\sqrt{\frac{t^2}{4+t^2}}, \quad t=\sqrt{\frac{4\tanh^2{r}}{1-\tanh^2{r}}} 
\end{equation}

One can plug the equation \ref{eq:cubic_to_quadratic} to the equation \ref{eq:squeezing_to_quadratic}, therefore, up to a global phase, obtains a relation between squeezing operation and concatenation of phase shifting, displacements and cubic phase gate operations:

\begin{eqnarray}
e^\frac{r(\hat{a}^2-\hat{a}^{\dag^2})}{2}=e^{-i\phi_1\hat{a}^\dag\hat{a}} e^{it_1P}e^{i\frac{1}{3}t_2X^3}e^{-2it_1P}\nonumber\\
\times e^{-i\frac{1}{3}t_2X^3}e^{it_1P}e^{-i\phi_2\hat{a}^\dag\hat{a}}\label{eq:final_relation}
\end{eqnarray}

where the product $t_1t_2$ is equivalent to $t$.


The relation between the squeezing parameter $r$ and the decibel of squeezing is as follows:

\begin{equation}
-10\log_{10}{e^{-2r}}=squeezing(\text{db})
\end{equation}

For example 10db squeezing corresponds to around $r=1.15$, thus, to obtain a 10db squeezed state the product $t_1t_2$ in eq. (\ref{eq:final_relation}) should be equivalent to $2.8416$. One can obtain this value through different combinations of $t_1$ and $t_2$. One would want to use small parameter for the cubic phase gate while it would be hard to implement this gate. Let's assume $t_2=0.1$ which gives an effective cubic phase gate interaction around $0.0333$ while the displacement would be in this case around $28.416$. This is not a problem because displacement operations are the easiest operations in the optical toolbox alongside phase shifting operations. Same logic applies for generating higher decibel squeezing operations. For more squeezed states one can just increase the displacement amplitudes while keeping the cubic phase amplitude same or even lower.


%

\section{Photon counting}

In this section we will present a method to measure photon statistics using the cubic phase gate.

Assume that we have given the following unitary evolution: 

\begin{equation}\label{eq:photon_counting_interaction}
e^{i\theta(X_1^2+P_1^2)X_2}
\end{equation}

where $\theta$ is the interaction time, and the subscripts denote the modes. Now we apply this operation to a two mode state $\sum c_n\ket{n}_1\ket{p\approx0}_2$ where the first mode is an arbitrary normalized state represented with Fock space expansion and the second mode is in approximate quadrature $P$ eigenstate with the eigenvalue 0:

\begin{equation}\label{eq:p_n_transformation}
e^{i\theta(X_1^2+P_1^2)X_2}\sum_n c_n\ket{n}_1\ket{p\approx0}_2 = \sum_n c_n\ket{n}_1\ket{p\approx \frac{\theta}{2} (n+\frac{1}{2})}_2
\end{equation}

Thus, there is a shift in the second mode which depends on the photon number in the first mode. Therefore, after this interaction, making a homodyne measurement in the second mode will correspond to a quantum non-demolition photon number detection in the first mode. In contrast to the Kerr based photon counting ideas this is not a rotation in phase space and can be discriminated with homodyne detection if the second mode is initially squeezed enough or the interaction parameter $\theta$ is big enough. We will deal with the problem of discrimination later, let us first discuss how to realize this interaction experimentally.

The experimental toolbox we have consists of Gaussian operations plus the cubic phase gate. First of all we split the operator in the following way:

\begin{equation}\label{eq:splitting}
e^{i\theta(X_1^2+P_1^2)X_2}\approx e^{i\theta X_1^2X_2}e^{i\theta P_1^2X_2}+O(\theta^2)
\end{equation}

Note that this approximation is valid for $\theta \ll 1$. One can also use better splitting approximations such as Suzuki-Yoshida \cite{suzuki1990fractal}, \cite{yoshida1990construction} splittings. Now the problem is reduced to the implementing the operators $e^{i\theta X_1^2X_2}$ and $e^{i\theta P_1^2X_2} $ 

It is possible to exactly decompose the interaction $e^{i\theta X_1^2X_2} $ to two cubic phase gate and Gaussian operations. Consider the following relations:  

\begin{eqnarray*}
 e^{it_1 P_1X_2} e^{it_2X_1^3} e^{-it_1 P_1X_2}=e^{it_2X_1^3+\frac{3}{2}it_1t_2X_1^2X_2+\frac{3}{4}it_1^2t_2X_1X_2^2+\frac{1}{4}it_1^3X_2^3}\\
  e^{-it_1 P_1X_2} e^{-it_2X_1^3} e^{it_1 P_1X_2}=e^{-it_2X_1^3+\frac{3}{2}it_1t_2X_1^2X_2-\frac{3}{4}it_1^2t_2X_1X_2^2-\frac{1}{4}it_1^3X_2^3}
 \end{eqnarray*}
  
  when we combine these two equations:
  
  \begin{equation}
e^{it_1t_2X_1^2X_2} = e^{it_1 P_1X_2} e^{i\frac{t_2}{3}X_1^3} e^{-2it_1 P_1X_2} e^{-i\frac{t_2}{3}X_1^3} e^{it_1 P_1X_2}
  \end{equation}

Similarly 

  \begin{equation}
e^{it_1t_2P_1^2X_2} = e^{-it_1 X_1X_2} e^{i\frac{t_2}{3}P_1^3} e^{2it_1 X_1X_2} e^{-i\frac{t_2}{3}P_1^3} e^{-it_1 X_1X_2}
  \end{equation}

Thus we get a set of interactions which consists of cubic phase gate and Gaussian operations.


We want the proposed setup to be photon number resolving. This means that, after the transformations (\ref{eq:p_n_transformation}), for each photon number second mode has to be distinguishable:

\begin{equation}
\braket{p\approx \frac{\theta}{2} (n'+\frac{1}{2})|p\approx \frac{\theta}{2} (n+\frac{1}{2})}_2 \approx \delta_{n,n'}
\end{equation}

It is sufficient to only consider the single photon differences. A single photon difference will shift the second finite squeezed mode in phase space with $d=\frac{3\theta}{4}$:

\begin{equation}
\left(\frac{2e^{2 r}}{\pi }\right)^{1/4}\int e^{-e^{2 r} p^2}\ket{p}dp \rightarrow \left(\frac{2e^{2 r}}{\pi }\right)^{1/4} \int e^{-e^{2 r}(p-d )^2}\ket{p}dp
\end{equation}

The overlap of a finite squeezed vacuum and displaced squeezed vacuum state is:

\begin{equation}
 \sqrt{\frac{2e^{2 r}}{\pi }} 	\int e^{-e^{2 r} p^2}e^{-e^{2 r} (p-d )^2}dp=e^{-\frac{1}{2} d ^2 e^{2 r}}
\end{equation}

If we want the presented setup photon number discriminating, than this overlap must be sufficiently small. Let's assume that the overlap should be smaller than or equal to $10^{-2}$. This leads to the following relation: $e^{r}*d\leq 3.03485$. 

One way to satisfy this constraint is to apply decomposition of the interaction (\ref{eq:photon_counting_interaction}) many times. This will effectively increase $\theta$. A usual 10db squeezing will correspond to $r=1.15$, thus, for $\theta \geq 1.28$ the constraint will be satisfied. Therefore if we decompose the interaction (\ref{eq:photon_counting_interaction}) for $\theta=0.1$ we have to apply the decomposition around 13 times.

Another way is to generate extreme squeezed states. While keeping the interaction parameter $\theta =0.1$ necessary squeezing will be 32db. This is possible with the squeezing method that we presented above. Or one can just combine these two ideas.
 
 To check the quality of the decomposition we made some numerical calculations. We applied the decomposition on to a input states where the first mode is in two photon state and the second mode is a 10db finite squeezed state. We used the initial interaction time of $\theta=10^{-1}$. We applied the decomposition on to this particular input state for 14 times, then we traced over the first mode to find the second mode. We evaluated the fidelities of the output state with the states: $\ket{p\approx \frac{\theta}{2} (1+\frac{1}{2})}$, $\ket{p\approx \frac{\theta}{2} (2+\frac{1}{2})}$, $\ket{p\approx \frac{\theta}{2} (3+\frac{1}{2})}$ . We used the usual fidelity measure for two matrices $A$ and $B$: $F(A,B)=\text{Tr}(AB)$. we found the following results: 

\begin{eqnarray*}
F(output,|p\approx \frac{\theta}{2} (1+\frac{1}{2})><p\approx \frac{\theta}{2} (1+\frac{1}{2})|)=0.0085\\
F(output,|p\approx \frac{\theta}{2} (2+\frac{1}{2})><p\approx \frac{\theta}{2} (2+\frac{1}{2})|)=0.9886\\
F(output,|p\approx \frac{\theta}{2} (3+\frac{1}{2})><p\approx \frac{\theta}{2} (3+\frac{1}{2})|)=0.0080
\end{eqnarray*}
%

 The results confirm our prediction, thus we can conclude that the decomposition is working properly. Note that instead of 13 we run it for 14 times because of the error induced by the splitting approximations in eq. (\ref{eq:splitting}) and numerical errors.

 \section{Improving the MFF method}
 
 Up to now we made two assumptions about the cubic phase gate. We assumed that it is a low amplitude gate. This is an important assumption while all the two proposals for the cubic phase gate is for small amplitude interaction. Our second assumption about the cubic phase gate was that it is perfect. This is not satisfied by the current experimental status, however one may also not need a perfect cubic phase gate to realize our proposals. In this section we review our proposals in terms of the Marek, Filip, Furusawa method \cite{Marek2011} make some proposals to improve it. MFF method is the experimentally realized method \cite{Yukawa2013}.
 
 The cubic phase state is not normalizable. In the MFF method, this manifest itself through finite squeezing as a Gaussian envelope: 
  
  \begin{equation}
\mathcal{N} \int  e^{itX^3}e^{-\frac{x^2}{e^{2r}}}\ket{x}dx
  \end{equation}
  
 Also, MFF method is a first order approximation. This cause it to be well approximated around the center in the phase space but away from the center it becomes useless. Both of these problems are ignorable depending on the input state. Because the input states are always normalized and in general, localized around a certain region in the phase space. Thus, to minimize the errors that arise from the Gaussian envelope and low order approximation, one can generate the cubic phase state that is less distorted and better approximated around the region in the phase space that the input is localized. This is possible. For example in our squeezing proposal, to produce squeezed vacuum, first a displacement operation is applied to a vacuum, this is a coherent state, and then a cubic phase gate is applied. Therefore, one needs to generate a cubic phase gate that is well approximated around the region that this coherent state localized. Gaussian envelope can be localized around this coherent state by creating the cubic phase gate from a displaced and finite squeezed state which is displaced to the region that the coherent state is localized. This leads to the following state: $\mathcal{N} \int e^{itX^3}e^{-\frac{(x-c)^2}{e^{2r}}}\ket{x}dx$ where $c$ is the amplitude for the initial displacement.
 
 Localized cubic gate idea can also be applied to the photon counting proposal. If the cubic phase gate is applied to weak input singles (states that contain few photons) than the current realization could be sufficient \cite{Yukawa2013}, in case one can obtain the necessary squeezing by other means.

%

 
 MFF method has been realized through three heralded photon detection \cite{Yukawa2013}. In principle it can be extended to higher orders. For example in case one needs a second order approximation than one needs six heralded photon detection. Thus the complexity of the setup and the probability of creating a proper state will increase exponentially with the order. Now, we show that when one has access to an optical memory, an arbitrary order can be decomposed to concatenations of the first order approximation, thus three heralded photon detections.
 
 Since Suzuki \cite{suzuki1990fractal}, it is known how to generate arbitrary higher order approximations recursively using first order approximations, thus combination of the first order approximations can be use to generate arbitrary higher order approximations. Consider an operator: $e^{A}$ and assume that you have $n$'th order Taylor approximation fo this particular operator. Obviously, following relation holds:
 
 \begin{equation}\label{eq:how_to_get_higher_orders}
 e^{A}=e^{cA}e^{(1-c)A}.
 \end{equation}
 
 Suzuki's way to do this is to eliminate the contribution from the $(n+1)$'th order terms on the right hand side to assure to obtain the $(n+1)$'th order term through cross products\footnote{Note that Suzuki used concatenation of three operators because he also imposed the condition that the solution set must be real, here we can relax this condition. However three concatenation can be also helpful to have a symmetrical decomposition which has nice properties. See the ref. \cite{suzuki1990fractal}.}. In addition to the condition $(c+1-c)=1$ which is obviously satisfied, this leads to the following condition:
 
 \begin{equation}
 c^{n+1}+(1-c)^{n+1}=0.
 \end{equation}
 
 It is also possible to match both sides for more than one term, for example for the orders $(n+1)$, $(n+2)$, etc.... For this purpose, more than two concatenations are needed in the right hand side of the equation \eqref{eq:how_to_get_higher_orders}, because there will be more than two equations to satisfy and things get more complicated because of the cross terms. Also a symmetrical concatenation will automatically 

\section{Summary}

In this work we explored two possibilities concerning the cubic phase gate. We proposed to use cubic phase state to implement squeezing operations and to use photon statistics experiments. We also proposed some improvements in the MFF method. Even though the current experimental realizations does not suffice to realize these proposals in the future with better experimental realizations these two proposals might be possible.
 
 \section{Acknowledgment}
 
 We thank to Peter van Loock for comments and discussions.


\end{document}